\documentclass[12pt,letterpaper]{article}
\usepackage{amsfonts}
\usepackage{amsmath}
\usepackage{geometry}
\usepackage[authoryear,round,longnamesfirst]{natbib}
\usepackage{xcolor}
\usepackage{grffile}
\usepackage{graphicx}

\newtheorem{theorem}{Theorem}
\newtheorem{corollary}{Corollary}

\newenvironment{proof}[1][Proof]{\noindent\textbf{#1.} }{\ \rule{0.5em}{0.5em}}
\allowdisplaybreaks

\begin{document}

\title{Extreme Changes in Changes\thanks{\setlength{\baselineskip}{4.5mm} We thank the editor, associate editor, two anonymous referees, Alfonso Flores-Lagunes, Hilary Hoynes, Doug Miller, and David Simon for useful advice about our empirical application. We benefited from discussions with Jon Roth. All
remaining errors are ours.
A Stata command, \texttt{ecic} (extreme changes in changes), associated with this paper can be installed from SSC archive with the following command line: \underline{\texttt{ssc} \texttt{install} \texttt{ecic}.}}}
\author{ Yuya Sasaki\thanks{%
Associate professor of economics, Vanderbilt University. Email:
yuya.sasaki@vanderbilt.edu} \ and \ Yulong Wang\thanks{%
Assistant professor of economics, Syracuse University. Email:
ywang402@syr.edu.}}
\date{}
\maketitle

\begin{abstract}
\setlength{\baselineskip}{8.25mm} Policy analysts are often interested in
treating the units with extreme outcomes, such as infants with extremely low
birth weights. Existing changes-in-changes (CIC) estimators are tailored to
middle quantiles and do not work well for such subpopulations. This paper
proposes a new CIC estimator to accurately estimate treatment effects at
extreme quantiles. With its asymptotic normality, we also propose a method
of statistical inference, which is simple to implement. Based on simulation
studies, we propose to use our extreme CIC estimator for extreme, such as
below 5\% and above 95\%, quantiles, while the conventional CIC estimator
should be used for intermediate quantiles. Applying the proposed method, we
study the effects of income gains from the 1993 EITC reform on infant birth
weights for those in the most critical conditions.
This paper is accompanied by a Stata command.

{\small {\ \ \ \newline
} \newline
}
\end{abstract}

\newpage 

\section{Introduction}\label{sec:introduction}


The difference-in-differences (DID) approach is a widely employed empirical
strategy for program evaluation in the presence of policy events in time.
The common DID methods critically depend on parallel trend assumptions and
focus on identifying (conditional) average effects. An alternative empirical
strategy is the changes in changes (CIC) method proposed by %
\citet{athey2006identification}. At the cost of alternative distributional
assumptions, the CIC gets around the common trend assumption and further can
identify distributions of counterfactual outcomes as opposed to just their
averages. Thus, CIC can be used to analyze heterogeneous individuals via
quantile treatment effects under rank invariance.

As are the cases with other quantile-based estimands, however, the existing CIC estimator only works for intermediate quantiles in theory. 
Practically, for instance, such an estimator is accurate for intermediate quantile levels such as $q \in (0.05,0.95)$ between the fifth to the ninety-fifth percentiles. 
This limitation for the existing CIC estimator rules out causal inference for those individuals at the extreme top and extreme bottom quantiles. 
Yet, it is sometimes rather at extreme quantiles that treatment effects are more relevant to social policy analysis. 
For instance, policymakers often care about treating economically disadvantaged subpopulations like the poorest individuals characterized by the limit $q \rightarrow 0$. 
The treatment effect for these tail subpopulations could be substantially larger than that for the mid-sample subpopulations, and hence it is imperative for such policymakers to have methods with which they can accurately assess treatment effects at the subpopulations in the tail.

In this paper, we propose an alternative CIC estimator that more accurately estimates the treatment effects at the tails, technically in the limits as $q \rightarrow 0$ and $q \rightarrow 1$. 
We also develop asymptotic normality for this estimator and propose an easy-to-construct confidence interval.
Based on our simulation studies, we provide the following practical recommendation. 
For the intermediate quantiles, use the existing estimator by \citet{athey2006identification} along with its standard error. 
For the extreme quantiles, on the other hand, use our proposed estimator along with its standard error. 
We suggest using the log-log plot to choose the switching point and demonstrate a combined use of both estimators in our empirical application.

With the proposed econometric method, we revisit the study by %
\citet{hoynes2015income} in which they use the 1993 event of EITC reform to
evaluate the effects of income gains on infant birth weights. While they
analyze average effects via the DID, we focus on the effects at the low
quantiles to see if such income gains can improve infant birth weights,
particularly for those at the most critical birth weight conditions. This
empirical question is of interest because low infant birth weight is known
to have long-lasting impacts on the health and economic well-being in
adulthood \citep[e.g.,][]{currie2011inequality} as well as an immediate
impact on infant mortality.

\noindent \textbf{Literature.} In contrast to the nowadays extensive body of
literature on DID, the literature on CIC is relatively thin. Since its first
proposal by \citet{athey2006identification}, the CIC framework has been
extended to fuzzy treatment assignments \citep{de2014fuzzy}, models with
covariates \citep{melly2015changes}, continuous treatments %
\citep{dhaultoeuille2022nonparametric}, and correction of attrition bias %
\citep{ghanem2022correcting}. To our best knowledge, however, no preceding
paper investigates extreme quantiles in the context of CIC, despite the
aforementioned policy relevance. On the other hand, there are a few papers
that investigate treatment effects at extreme quantiles outside the context
of CIC -- see \citet{chernozhukov2005extremal}, %
\citet{chernozhukov2011inference}, \citet{d2018extremal}, %
\citet{zhang2018extremal}, and \citet{deuber2021estimation} to list but a
few. None of the existing papers on extremal treatment effects consider DID
or CIC frameworks.

\noindent \textbf{Organization.} Section \ref{sec:review} provides a review
of CIC. Section \ref{sec:ecic} introduces the proposed method, and Section %
\ref{sec:theory} derives asymptotic properties. 
Section \ref{sec:practice} discusses some practical issues, and Section \ref{sec:covariates} extends the proposed method to allow for covariates. 
Section \ref{sec:simulations} shows simulation studies, and Section \ref{sec:application} presents the
empirical application. Section \ref{sec:simulations_empirical} presents additional simulation results calibrated to the empirical dataset, and Section \ref{sec:conclusion} concludes.

\noindent \textbf{Stata Command.} This paper is accompanied by a Stata command, \texttt{ecic} (extreme changes in changes).
The package can be installed from SSC archive with the following command line: \underline{\texttt{ssc} \texttt{install} \texttt{ecic}}.
After the installation, run \underline{\texttt{help} \texttt{ecic}} for usage of the command.


\section{The Changes in Changes}

\label{sec:review} 
This section briefly reviews the CIC estimator following \citet{athey2006identification}. 
The goals here are to introduce the data-generating model and the treatment parameter of interest, as well as to fix notations to be used in the rest of this paper.

Individual $i$ belongs to group $G^{i}\in \{0,1\}$, where value of 0
(respectively, 1) indicates the control (respectively, treatment) group.
Each individual is observed in one of the two time periods $T^{i}\in \{0,1\}$%
. For each draw $i=1,...,n$ from the population, the group identity $G^i$
and time period $T^i $ are treated as random variables. Letting $Y^{i}$
denote a continuous outcome, econometricians observe a random sample of $%
(Y^{i},G^{i},T^{i})$.

The underlying structure to generate $Y^i$ is as follows. Let $Y^{i}_{N}$
(respectively, $Y^{i}_{I}$) denote the potential outcome for individual $i$
under no treatment (respectively, under treatment). The potential outcome $%
Y_N$ under no treatment is generated by 
\begin{align}
Y_{N}=h\left( U,T\right) ,  \label{eq:model}
\end{align}
where $U$ represents unobserved characteristics, $h\left(\cdot,t\right) $ is
strictly increasing for each $t\in \{0,1\}$, and $U \perp T|G$. Let $%
I^{i}\in \{0,1\}$ indicate that individual $i$ receives a treatment. In the
two-group-two-period setting, we have $I^{i} = G^{i} T^{i}$. The realized
outcome $Y^{i}$ is generated by 
\begin{align*}
Y^{i}=Y^{i}_{N}\left( 1-I^{i}\right) +Y^{i}_{I}\cdot I^{i}.
\end{align*}

We now introduce the following short-hand notations: 
\begin{align*}
Y_{gt}^{N} &\sim Y_{N}|G=g,T=t \\
Y_{gt}^{I} &\sim Y_{I}|G=g,T=t \\
Y_{gt} &\sim Y|G=g,T=t.
\end{align*}
For any distribution function $F$, we define its left-inverse $F^{-1}$ by $%
F^{-1}\left( q\right) =\inf \{y:F \left( y\right) \geq q\}$. In this setup
and with these notations, \citet[][Theorem 3.1]{athey2006identification}
establish 
\begin{equation*}
F_{Y_{11}^{N}}\left( y\right) =F_{Y_{10}} \circ F_{Y_{00}}^{-1} \circ
F_{Y_{01}}\left( y\right)
\end{equation*}
for all $y$, provided that $U|G=1$ is a subset of the support of $U|G=0$.

For each quantile $q\in (0,1)$, the quantile effect of the treatment is thus
identified by 
\begin{equation}
\tau _{q}^{CIC} := F_{Y_{11}^{I}}^{-1}\left(
q\right)-F_{Y_{11}^{N}}^{-1}\left( q\right) = F_{Y_{11}}^{-1}\left(
q\right)-F_{Y_{01}}^{-1} \circ F_{Y_{00}} \circ F_{Y_{10}}^{-1}\left(
q\right).  \label{eq:cic}
\end{equation}


\section{The Extreme Changes in Changes}

\label{sec:ecic} 
The conventional estimator \citep[][page 464]{athey2006identification} for $%
\tau_q^{CIC}$ performs well in middle quantiles, such as $q \in (0.05, 0.95)$%
, but may perform less desirably in extreme quantiles (e.g., $q \in
(0.00,0.05] \cup [0.95,1.00)$), as are the case with common quantile
estimators. Indeed, the asymptotic theory for the conventional estimator
rules out extreme values of $q$. In this section, we present our proposed
method of estimating $\tau^{CIC}_q$ as $q = q_n \rightarrow 1$ in the right
tail. A symmetric argument applies to the limit on the other side of the
distribution in the left tail as $q \rightarrow 0$. To stress the drifting
sequence of limiting parameters of our interest, we use the notation $\tau
_{q}^{eCIC}$ for extreme CIC. In other words, `e' in ``eCIC'' is used to
remind readers that this parameter drifts with $q$ as the sample size
increases.

Suppose that the distribution function $F_{Y_{gt}}$ of $Y_{gt}$ has
regularly varying tails for each $\{g,t\}\in \{0,1\}^{2}$. Specifically, we
assume 
\begin{equation*}
\frac{1-F_{Y_{gt}}\left( ty\right) }{1-F_{Y_{gt}}\left( t\right) }
\rightarrow y^{-\alpha _{gt}}\text{ as }t\rightarrow \infty
\end{equation*}
for each $\{g,t\}\in \{0,1\}^{2}$. Here, the parameter $\alpha _{gt}>0$ is
referred to as the Pareto exponent. Our extreme CIC estimation is built on
estimating the Pareto exponent. We emphasize that this assumption is quite
mild and most of the common families of parametric distributions as well as
a large class of nonparametric distributions satisfy it. For example, the
Student-t distribution with $\nu$ degrees of freedom satisfies this
condition with $\nu$ being the Pareto exponent. See, for example, %
\citet[][Chapter 1]{de2007extreme} and \citet[][Chapter 2]{resnick2007book}
for reviews on this condition.

Let $Y_{gt}^{(1)}\geq Y_{gt}^{(2)}\geq ...\geq Y_{gt}^{(n_{gt})}$ denote the
order statistics of the realized outcomes in the group $\{g,t\}$, where $%
n_{gt}$ denotes the subsample size in this group. Choose the largest $%
k_{gt}+1$ of them, that is 
\begin{equation*}
Y_{gt}^{(1)}\geq Y_{gt}^{(2)}\geq ... \geq Y_{gt}^{(k_{gt}+1)}.
\end{equation*}
Then, $\alpha _{gt}$ can be estimated by the Hill estimator %
\citep{hill1975simple} 
\begin{equation}
\hat{\alpha}_{gt}=\left( \frac{1}{k_{gt}}\sum_{i=1}^{k_{gt}}\left[ \log
\left( Y_{gt}^{(i)}\right) -\log \left( Y_{gt}^{\left( k_{gt}+1\right)
}\right) \right] \right) ^{-1}.  \label{eq:hill}
\end{equation}
As $q \rightarrow 1$, $F_{Y_{gt}}^{-1}\left( q\right) $ is estimated by 
\begin{equation}
\hat{F}_{Y_{gt}}^{-1}\left( q\right) =Y_{gt}^{\left( k_{gt}+1\right) }\left( 
\frac{k_{gt}}{n_{gt}\left( 1-q\right) }\right) ^{1/\hat{\alpha} _{gt}}.
\label{eq:q_est}
\end{equation}
Moreover, the tail probability can be estimated by 
\begin{equation}
1-\hat{F}_{Y_{gt}}\left( y\right) =\frac{k_{gt}}{n_{gt}}\left( \frac{y}{
Y_{gt}^{(k_{gt}+1)}}\right) ^{-\hat{\alpha}_{gt}}  \label{eq:prob_est}
\end{equation}
as $q \rightarrow 1$. See, for example, \citet[][Chapter 4]{de2007extreme}.

By combining the identifying formula \eqref{eq:cic} with the component
estimators \eqref{eq:hill}--\eqref{eq:prob_est}, we obtain the following
estimator for the extreme CIC, $\tau _{q}^{eCIC}$. 
\begin{align*}
\hat{\tau}_{q}^{eCIC} =&\hat{F}_{Y_{11}}^{-1}\left(q\right)-\hat{F}_{Y_{01}}^{-1} \circ \hat{F}_{Y_{00}} \circ \hat{F}_{Y_{10}}^{-1}\left(q\right) \\
=&Y_{11}^{\left( k_{11}+1\right) }\left( \frac{k_{11}}{n_{11}\left( 1-q\right) }\right) ^{1/\hat{\alpha}_{11}} -Y_{01}^{\left( k_{01}+1\right) }\left( \frac{k_{01}}{n_{01} \left(1-\hat{F}_{Y_{00}} \circ \hat{F}_{Y_{10}}^{-1}\left(q\right)\right) }  \right) ^{1/\hat{\alpha}_{01}} \\
=&Y_{11}^{\left( k_{11}+1\right) }\left( \frac{k_{11}}{n_{11}\left( 1-q\right) }\right) ^{1/\hat{\alpha}_{11}} \\
&-Y_{01}^{\left( k_{01}+1\right) }\left( \frac{k_{01}}{n_{01}}\frac{n_{00}}{%
k_{00}}\left( \frac{\left( Y_{10}^{\left( k_{10}+1\right) }\left( \frac{
k_{10}}{n_{10}\left( 1-q\right) }\right) ^{1/\hat{\alpha}_{10}}\right) }{%
Y_{00}^{\left( k_{00}+1\right) }}\right) ^{\hat{\alpha}_{00}}\right) ^{1/\hat{\alpha}_{01}}.
\end{align*}
By simple algebraic manipulations, this expression simplifies as 
\begin{align}
\hat{\tau}_{q}^{eCIC} =&Y_{11}^{\left( k_{11}+1\right) }\left( \frac{k_{11}}{%
n_{11}}\right) ^{1/\hat{\alpha}_{11}}\left( 1-q\right) ^{-1/\hat{\alpha}%
_{11}}  \notag \\
&-Y_{01}^{\left( k_{01}+1\right) }\left( \frac{Y_{10}^{\left(k_{10}+1\right)
}}{Y_{00}^{\left( k_{00}+1\right) }}\right) ^{\hat{\alpha}_{00}/\hat{\alpha}%
_{01}}\left( \frac{k_{01}}{n_{01}}\frac{n_{00}}{k_{00}} \right) ^{1/\hat{%
\alpha}_{01}}  \label{eq:cic_est} \\
&\times \left( \frac{k_{10}}{n_{10}}\right) ^{\hat{\alpha}_{00}/\left( \hat{%
\alpha}_{10}\hat{\alpha}_{01}\right) }\left( 1-q\right) ^{-\hat{\alpha}%
_{00}/\left( \hat{\alpha}_{10}\hat{\alpha}_{01}\right) }.  \notag
\end{align}
We thus propose \eqref{eq:cic_est} as the extreme CIC estimator, which is
quite simple to implement.
The next section presents asymptotic properties based on $k_{gt}\rightarrow \infty$ as $n_{gt}\rightarrow \infty$ for all $g$ and $t$. 

We close this section with a discussion of the identifiability of the extreme CIC, $\tau^{eCIC}_q$.
While we informally reviewed the identification result of \citet[][Theorem 3.1]{athey2006identification} in Section \ref{sec:review}, we should emphasize that it relies on a common support condition \citep[][Assumption 3.4]{athey2006identification}.
Namely, for the identifying equality \eqref{eq:cic} to hold for \textit{all} $q \in (0,1)$, the support of $U|G=1$ needs to be a subset of the support of $U|G=0$.
If this condition fails, then $F_{Y^N_{11}}$ remains unidentified outside of the support of $Y|G=0,T=1$ \citep[][Corollary 3.1]{athey2006identification}.
Such an unidentified region of $q$ generally contains extreme quantiles.
Hence, the common support condition is crucial especially in the context of extreme quantiles.
If $F_{Y^N_{11}}$ is bounded away from zero and one on the support of $Y|G=0,T=1$, then we can deduce that the common support condition may be violated.


\section{Asymptotic Theory}

\label{sec:theory} 
In this section, we derive a limit distributional property for the proposed
extreme CIC estimator. This result paves a way for statistical inference
about the extreme CIC.

Let $\{Y_{gt}^{i}\}_{i=1}^{n_{gt}}$ denote the subsample of observed
outcomes in group $g$ and time $t$. We state the following set of
conditions, followed by discussions of each piece.

\bigskip\noindent \textbf{Conditions}
\begin{enumerate}
\item $Y_{gt}^{i}$ is i.i.d. across $i$ within each $g$ and $t$. $%
\{Y_{gt}^{1},...,Y_{gt}^{n_{gt}}\}$ are independent across $g$ and $t$.

\item $F_{gt}\left( \cdot \right) $ is regularly varying at infinity with
Pareto exponent $\alpha _{gt}$. Moreover, for some constant $\rho_{gt}>0$, $%
1-F_{gt}\left( y\right) =c_{1}y^{-\alpha _{gt}}+c_{2}y^{-\alpha
_{gt}-\alpha_{gt}\rho _{gt}}\left( 1+o(1)\right) \text{ as }y\rightarrow
\infty . $

\item $n_{11}/n_{gt}\rightarrow \eta _{11/gt}\in (0,\infty )$ and $%
k_{11}/k_{gt}\rightarrow \lambda _{11/gt}\in \left( 0,\infty \right) $ for
each $g,t\in \{0,1\}^{2}$.

\item $k_{gt}\rightarrow \infty $ and $k_{gt}=o\left( n_{gt}^{2\rho
_{gt}/\left( 1+2\rho _{gt}\right) }\right) $ for each $g,t\in \{0,1\}^{2}.$

\item $n_{gt}(1-q)=o(k_{gt})$ and $\log \left[ n_{gt}\left( 1-q\right) %
\right] =o\left( \sqrt{k_{gt}}\right) $ for each $g,t\in \{0,1\}^{2}.$

\item $F_{Y_{11}}^{-1}\left( q\right) /F_{Y_{01}}^{-1} \circ F_{Y_{00}}
\circ F_{Y_{10}}^{-1}\left( q\right) \rightarrow \varsigma \in \left(
0,\infty \right) $.
\end{enumerate}

\bigskip We provide some discussions about these conditions. 
Following \citet{athey2006identification}, Condition 1 assumes random sampling within each time and treatment group, and independence across time periods and groups. 
Thus, it presumes repeated cross sections rather than panel data. 
Condition 2 imposes the regularly varying tail conditions on all four conditional distributions of the outcome. 
More generally, the regularly varying tail condition is equivalent to that the underlying distribution belongs to the domain of attraction of the extreme value distribution with a positive tail index. 
See, for example, \citet[][Chapter 1]{de2007extreme}. 
Since we derive the convergence rate, the second-order Pareto tail approximation is inevitable. 
The second-order parameter $\rho _{gt}$ governs the distance between the true underlying distribution and the Pareto one. 
As remarked previously, this condition imposes a rather mild restriction and also satisfies the common support condition. 
Condition 3 requires that the sample sizes of all subsamples are asymptotically of the same order of magnitude. 

Condition 4 specifies the order of the tail thresholds used in estimation. 
For simplicity of illustration, we select $k_{gt}$ to be of a smaller order than $n_{gt}^{2\rho _{gt}/\left( 1+2\rho_{gt}\right) }$ so that the estimators incur negligible asymptotic biases relative to variances. 
This requirement is similar in spirit to under-smoothing bandwidths in kernel estimation or under-smoothing dimensions in sieve estimation. 
On the other hand, if we select $k_{gt}$ to be of the same order of $n_{gt}^{2\rho _{gt}/\left( 1+2\rho_{gt}\right) }$, the asymptotic bias becomes non-negligible, whose expression is complicated. 
In particular, the asymptotic bias involves the second-order parameter $\rho_{gt}$ \citep[e.g.,][Chapter 3]{de2007extreme}. 
Estimation of this parameter is challenging since it requires further restrictions on the underlying distribution \citep[e.g.,][]{cheng2001confidence, haeusler2007, CarpentierKim2014}, which are hard to interpret and hard to justify. 
Furthermore, such bias estimators entail slower rates of convergence.
Given these limitations, we focus on the asymptotics based on undersmoothing for a better statistical inference.  

Condition 5 imposes restrictions on the rate at which the quantile level $q$ under investigation tends to the unit in the drifting sequence. 
In particular, $q$ should tend to one sufficiently fast so that the quantile under investigation is extreme. 
Otherwise, the $q$ quantile is not in the tail and can be better estimated by the standard CIC method. 
This condition is also common in the extreme quantile literature \citep[e.g.,][Chapter 4]{de2007extreme}. 
Note that this condition allows for $n_{gt}(1-q)\rightarrow 0$. 
When this happens, the other part of this condition implicitly imposes a lower bound of $1-q$ and equivalently that the extrapolation cannot be pushed too far in the right tail \citep[e.g.,][Remark 4.3.4]{de2007extreme}. 
To see this, observe that the condition $\log \left[ n_{gt}\left( 1-q\right) \right] =o\left( \sqrt{k_{gt}}\right)$ implies $1-q>n^{-1}\exp(-\varepsilon\sqrt{k_{gt}})$ for each $\varepsilon>0$. 

Condition 6 requires that the limit of the counterfactual outcome ratio is finite as $q$ tends to the unit. 
For simplicity, we consider $\varsigma \in (0,\infty)$.
If $\varsigma$ is $0$ or $\infty$, however, the estimator ${\hat{F}_{Y_{gt}}^{-1}(\cdot) }$ has a different convergence rate across $g$ and $t$, and consequently, we could ignore the estimation error for some pairs of $g$ and $t$.

The following theorem establishes the asymptotic normality for the extreme
CIC estimator \eqref{eq:cic_est} under these conditions.

\begin{theorem}
\label{thm:cic} If Conditions 1-6 are satisfied, then 
\begin{equation*}
\frac{k_{11}^{1/2}}{F_{Y_{11}}^{-1}\left( q\right) \log d_{11}}\left( \hat{%
\tau}_{q}^{eCIC}-\tau _{q}^{eCIC}\right) \overset{d}{\rightarrow }\mathcal{N}%
\left( 0,\Omega \right)
\end{equation*}
holds, where $d_{gt} = k_{gt} / (n_{gt}(1-q))$ and 
\begin{equation*}
\Omega =\alpha _{11}^{-2}+\left( \frac{1}{\varsigma }\right) ^{2}\left(\frac{%
\lambda _{11/10}}{\eta _{11/10}}\right) ^{2}\left[ \lambda_{11/00}+\lambda
_{11/10}+\lambda _{11/01}\right] \frac{\alpha _{00}^{2}}{\alpha
_{10}^{2}\alpha _{01}^{2}}.
\end{equation*}
\end{theorem}

\noindent A proof is relegated to Appendix \ref{sec:thm:cic}.

In finite samples, the asymptotic variance can be estimated by substituting $%
\hat{\alpha}_{gt}$, $\hat{\lambda}_{11/gt}=k_{11}/k_{gt}$, and $\hat{%
\varsigma}=\hat{F}_{Y_{11}}^{-1}\left( q\right) /\hat{F}_{Y_{01}}^{-1} \circ 
\hat{F}_{Y_{00}} \circ \hat{F}_{Y_{10}}^{-1}\left( q\right)$ for $\alpha
_{gt}$, $\lambda _{11//gt}$, and $\varsigma $, respectively, in the formula
of the asymptotic variance $\Omega $ provided in the statement of Theorem %
\ref{thm:cic}. Under the same conditions, this estimator of $\Omega $ is also consistent. 
The 95\% confidence interval is then constructed as 
\begin{equation}
\hat{\tau}_{q}^{eCIC}\pm 1.96k_{11}^{-1/2}\log d_{11}\left[ 
\begin{array}{c}
\hat{F}_{Y_{11}}^{-1}\left( q\right) ^{2}\hat{\alpha}_{11}^{-2}+\left( \hat{F%
}_{Y_{01}}^{-1} \circ \hat{F}_{Y_{00}} \circ \hat{F}_{Y_{10}}^{-1}\left(q%
\right) \right) ^{2} \\ 
\times \left( \frac{\hat{\lambda}_{11/10}}{\hat{\eta}_{11/10}}\right) ^{2}%
\left[ \hat{\lambda}_{11/00}+\hat{\lambda}_{11/10}+\hat{\lambda}_{11/01}%
\right] \frac{\hat{\alpha}_{00}^{2}}{\hat{\alpha}_{10}^{2}\hat{\alpha}%
_{01}^{2}}%
\end{array}
\right] ^{1/2}.  \label{eq:stderr}
\end{equation}

In practice, it is recommended to replace $d_{11}$ by $d_{11} \vee 
\underline{d}$ for some $\underline{d} > 1$ to ensure a positive value of
the logarithm in \eqref{eq:stderr}. We set $\underline{d} = 10$ in the
subsequent simulation studies and empirical application. Finally, we remark
that $\Omega $ simplifies to 
\begin{equation*}
\Omega =\alpha ^{-2}\left[1 + \left( \frac{1}{\varsigma }\right) ^{2}\left( 
\frac{\lambda _{11/10}}{\eta _{11/10}}\right) ^{2}\left[ \lambda
_{11/00}+\lambda _{11/10}+\lambda _{11/01}\right] \right]
\end{equation*}
in the special case where $\alpha _{g,t}$ is the same across $\{g,t\}$, say $%
\alpha_{g,t}=\alpha$ for all $\{g,t\}$, although we do not impose this
restriction in the subsequent numerical analyses. This could happen if the
treatment effect is a constant shift of the outcome. Given that $\hat{\alpha}%
_{g,t}$ is asymptotically independent across $g$ and $t$, we can perform the
standard t-test for their equivalence.


\section{Practical Issues}
\label{sec:practice}
This section collects discussions on the remaining practical issues.

\subsection{Choice of $k_{gt}$}\label{sec:kng}

The number $k_{gt}$ of order statistics is the key tuning parameter in our method.
We propose to use the  empirical choice rule proposed by \citet{guillou2001diagnostic}. 
We present the detailed procedure here for convenience of readers. 

Since the identical algorithm applies to each pair of $g$ and $t$, we suppress these subscripts in this subsection for notational simplicity. 
Given a random sample $\{Y^1,Y^2,\ldots,Y^n\}$, we first sort them descendingly and denote the order statistics by $Y^{(1)}\geq Y^{(2)} \geq \ldots \geq Y^{(n)}$. 
Define $Z_{i}=i\log(Y^{(i)}/Y^{(i+1)})$ for $i=1,\ldots ,n-1$. 
For each $k=1,\ldots, n-1$, construct
\begin{equation*}
\mathcal{T}_{k}\equiv \left( \sum_{i=1}^{k}w_{i}^{2}\right) ^{-1/2}\hat{\xi}^{-1}U_{k},
\end{equation*}
where $w_{i}=\text{sgn}\left( k-2i+1\right)\left\vert k-2i+1\right\vert $, $\hat{\xi} = 1/\hat{\alpha}$, and $U_{k}\equiv\sum_{i=1}^{k}w_{i}Z_{i}$. 
When the Pareto tail approximation performs well, $\mathcal{T}_{k}$ should have its mean close to zero and variance close to one. 
Accordingly, we can minimize the following criteria based on a moving average of $\mathcal{T}_{k}^{2}$: 
\begin{equation*}
\mathcal{C}_{k}=\left( \left( 2\lfloor k/2 \rfloor+1\right) ^{-1}\sum_{j=-l}^{\lfloor k/2 \rfloor}\mathcal{T}
_{k+j}^{2}\right) ^{1/2}.
\end{equation*}
The optimal value $k^\ast$ of $k$ is
\begin{equation}
k^{\ast }=\min_{1\leq k\leq n-1}\{k:\mathcal{C}_{t}>1\text{ for all }t\geq k\}.  
\label{k choice}
\end{equation}

\subsection{Extreme Quantiles}\label{sec:switching}

We now discuss how to define the domain $[\underline q, 1)$ of $q$ on which one may use this extreme CIC estimator, as opposed to the conventional CIC estimator.
We suggest to make a scatter plot of $\{\log Y_{gt}^{(i)}\}_{i=1}^{n_{gt}}$ against $\{\log i\}_{i=1}^{n_{gt}}$, called the log-log plot.
This plot is linear near small values of $i$ if the tail is approximately Pareto, and our estimator is accurate where it appears linear.
In this light, one can choose the boundary point $\underline q$ such that this log-log plot appears linear for $i \in \{1,\cdots,\lfloor n_{gt}(1 - \underline q) \rfloor\}$.
We concretely illustrate this procedure in our empirical application in Section \ref{sec:application}.

\section{Extension: Covariates}\label{sec:covariates}

Our proposed method can be easily extended to allow for covariates. Similarly to \citet[][pages 465-466]{athey2006identification}, we first regress the outcome variable on the covariates and then apply the proposed extreme CIC estimator to the regression residuals. We formalize this procedure as follows. 

Consider the linear model 
\begin{equation}
W_{gt}^{i}=\left( X_{gt}^{i}\right) ^{\prime }\beta _{gt}+Y_{gt}^{i},
\label{eq:ols}
\end{equation}%
where $W_{gt}^{i}$ denotes the outcome variable for the $i$-th individual in group $g$ and time $t$, and $X_{gt}^{i}$ denotes the covariate vector. 
The coefficient $\beta _{gt}$ can be different across $g$ and $t$, and hence the above regression can be conducted separately for each $g$ and $t$. 
For notational simplicity, we continue using $Y_{gt}^{i}$ to denote the error term, which is now unobserved. 
Given an estimate $\hat{\beta}_{gt}$, we treat the residuals $\hat{Y}_{gt}^{i}=W_{gt}^{i}-\left( X_{gt}^{i}\right)^{\prime }\hat \beta _{gt}$ as effective observations and construct the proposed extreme CIC estimator based on them. 
Specifically, we order the residuals as
\begin{equation*}
\hat{Y}_{gt}^{(1)}\geq \hat{Y}_{gt}^{(2)}\geq ...\geq \hat{Y}_{gt}^{(k_{gt}+1)}
\end{equation*}%
and replace $\{Y_{gt}^{(i)}\}_{i=1}^{k_{gt}+1}$ with $\{\hat{Y}_{gt}^{(i)}\}_{j=1}^{k_{gt}+1}$ in \eqref{eq:hill}--\eqref{eq:cic_est}.

In additional to Conditions 1-6, we require the following additional condition. 

\bigskip
\noindent \textbf{Condition}
\begin{enumerate}
\item[7.]
$\max_{1\leq i\leq n_{gt}}\sqrt{k_{gt}}\frac{\left\vert \hat{Y}_{gt}^{i}-Y_{gt}^{i}\right\vert }{1+\left\vert Y_{gt}^{i}\right\vert }=o_{p}(1)$ for all $g$ and $t$. 
\end{enumerate}

Condition 7 is proposed recently by \citet{girard2021}, who study an
estimator of tail features in a more general setup.  
This condition is mild and satisfied by the least square estimator in the linear model \eqref{eq:ols} \citep[cf.,][Section 3.1]{girard2021}. 
In particular, when $X_{gt}^{i}$ has a compact support, and the regression estimator $\hat{\beta}%
_{gt}$ is $\sqrt{n}$-consistent, $\vert \hat{Y}_{gt}^{i}-Y_{gt}^{i}\vert $ becomes $\vert\vert X_{gt}^{i}\vert\vert \cdot \vert\vert \hat{\beta}_{gt}-\beta _{gt} \vert \vert = O_p(\sqrt{n_{gt}})$ . Then
Condition 7 follows from that $k_{gt}/n_{gt}\rightarrow 0$. In summary, this
condition requires that the estimation error is sufficiently small and
consequently the CIC estimator based on $\{\hat{Y}_{gt}^{(i)}%
\}_{i=1}^{k_{gt}+1}$ is asymptotically the same as that based on $%
\{Y_{gt}^{(i)}\}_{i=1}^{k_{gt}+1}$. 

The following corollary summerzies the result.

\begin{corollary}
\label{col:cic} Consider the linear regresssion model \eqref{eq:ols}. If
Conditions 1-7 are satisfied, then the estimator $\hat{\tau}_{q}^{eCIC}$
based on $\{\hat{Y}_{gt}^{(i)}\}_{i=1}^{k_{gt}+1}$ has the same asymptotic
distribution as in Theorem 1.
\end{corollary}

\noindent A proof is relegated to Appendix \ref{sec:thm:cic}.

\section{Simulations}

\label{sec:simulations} 
We use the following data generating design based on our baseline model.
Generated first are the group and time period indicators according to 
\begin{align*}
G^{i} \sim \text{Bernoulli}(\pi_G) \quad\text{and}\quad T^{i} \sim \text{%
Bernoulli}(\pi_T).
\end{align*}
To allow for the endogenous dependence between the group $G^{i}$ and the
unobservables $U^{i}$, we in turn generate $U^{i}$ conditionally on $G^{i}$
as follows. 
\begin{align*}
U^{i} \sim 
\begin{cases}
\text{Beta}(\pi_A,\pi_B) & \text{if } G^{i}=0 \\ 
\text{Uniform}(0,1) & \text{if } G^{i} = 1.%
\end{cases}%
\end{align*}
Here, we use the uniform distribution under $G^{i}=1$ for ease of analytic
tractability of both the Pareto exponent and the quantile treatment effects
and for the purpose of accurate evaluations of simulation results with
analytically known true parameter values. We also remark that the
conditional independence assumption $U^{i} \perp T^{i}|G^{i}$ of %
\citet{athey2006identification} is satisfied in this design by construction.
In this two-group-two-period setting, the treatment indicator is in turn
defined by $I^{i} = G^{i} T^{i}$.

The potential outcomes are generated through the model 
\begin{align}
Y^{i}_N =& h^N(U^{i},T^{i}) = F_{t_{\alpha}}^{-1}(U^{i}) + T^{i} \qquad\text{%
and}  \label{eq:sim:YN} \\
Y^{i}_I =& h^I(U^{i},T^{i}) = F_{t_{\alpha}}^{-1}(U^{i}) + U^{i} + 1,
\label{eq:sim:YI}
\end{align}
where $F_{t_{\alpha}}^{-1}$ denotes the quantile function of the Student-t
distribution with $\alpha$ degrees of freedom. Now, the observed outcomes
are in turn generated by 
\begin{equation*}
Y^{i}=Y^{i}_{N}\left( 1-I^{i}\right) +Y^{i}_{I}\cdot I^{i}\text{.}
\end{equation*}

There are three notable features in this data generating process. First, $%
F_{Y^N_{11}}$ and $F_{Y^I_{11}}$ in \eqref{eq:sim:YN}--\eqref{eq:sim:YI}
have Pareto exponents of $\alpha$. 
Second, the second term $U$
on the right-hand side of \eqref{eq:sim:YI}, but not of \eqref{eq:sim:YN},
causes heterogeneous treatment effects characterized as follows 
\begin{align*}
\tau_q^{CIC} = F_{Y^I_{11}}^{-1}(q) - F_{Y^N_{11}}^{-1}(q) = q.
\end{align*}
Finally, we remark that the monotonicity assumption of %
\citet{athey2006identification} for the identification is satisfied in this
model.

We evaluate the finite sample performance of our proposed extreme CIC
estimator $\hat{\tau}_q^{eCIC}$ given in \eqref{eq:cic_est} along with its
standard error estimator \eqref{eq:stderr}. The order statistics $k_{gt}$
are chosen based on \citet{guillou2001diagnostic} for each subsample $(g,t)$ as described in Section \ref{sec:kng}. 
We also present simulation results for the conventional
estimator $\hat{\tau}_q^{CIC}$ of \citet[][page 464]{athey2006identification}
with its standard error estimator \citep[][pages
464-465]{athey2006identification}. For the standard error estimation for $%
\hat\tau_q^{CIC}$, we use Epanechnikov kernel and Silverman's rule of thumb
for bandwidth selection. Before presenting the results, we want to stress
that we focus on the extreme quantiles $q \in [0.90,1.00)$ on which
comparisons are necessarily unfair for the estimator $\hat{\tau}_q^{CIC}$ of %
\citet{athey2006identification}, which presumes intermediate quantiles in
theory. We confirm and acknowledge that the estimator $\hat{\tau}_q^{CIC}$
of \citet{athey2006identification} performs better in the intermediate
quantiles $q \in (0.10,0.90)$.

\begin{figure}[tbp]
\centering
\begin{tabular}{cc}
\includegraphics[width=6.6cm]{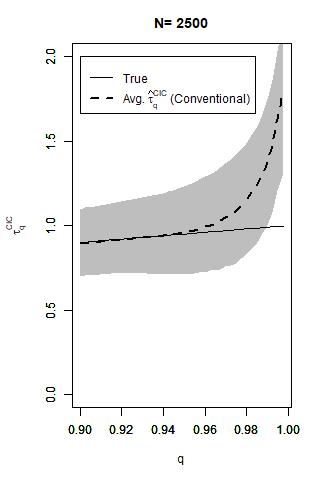}
& %
\includegraphics[width=6.6cm]{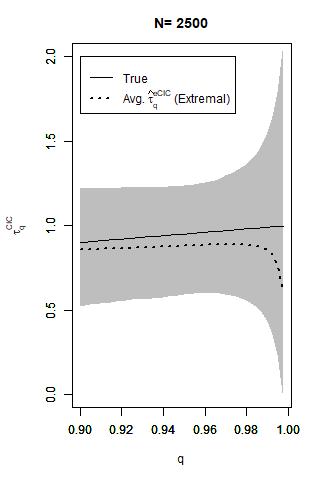}
\\ 
\includegraphics[width=6.6cm]{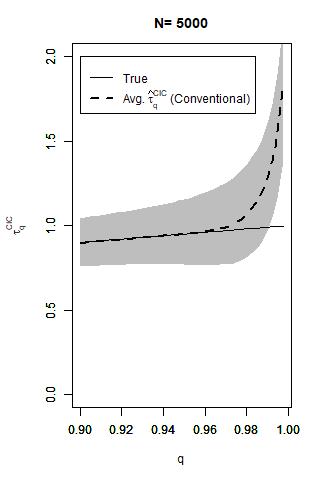}
& %
\includegraphics[width=6.6cm]{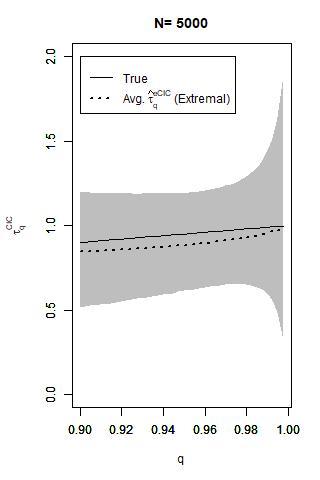}%
\end{tabular}
${}$%
\caption{\setlength{\baselineskip}{5.5mm}Monte Carlo averages and
inter-quartile ranges (shaded) of the estimates based on the conventional
estimator $\hat{\protect\tau}_q^{CIC}$ (dashed curves on the left column)
and our proposed estimator $\hat{\protect\tau}_q^{eCIC}$ (dotted curves on
the right column) at the extreme quantiles $q \in [0.90, 1.00)$ under the
design with $(\protect\pi_G,\protect\pi_T,\protect\pi_A,\protect\pi_B,%
\protect\alpha)=(0.1,0.5,1.0,2.0,10)$. The true treatment effects are
indicated by the solid curves.}
\label{fig:sim:estimates}
\end{figure}

Figure \ref{fig:sim:estimates} shows Monte Carlo averages and inter-quartile
ranges of the estimates under the design with $(\pi_G,\pi_T,\pi_A,\pi_B,%
\alpha)=(0.1,0.5,1.0,2.0,10)$. The dashed curves on the left column of the
figure indicate the average estimates based on the conventional estimator $%
\hat{\tau}_q^{CIC}$. The dotted curves on the right column of the figure
indicate the average estimates based on our proposed estimator $\hat{\tau}%
_q^{eCIC}$. In each panel, the shaded regions indicate the inter-quartile
ranges of the estimates by the respective methods. The results are shown at
the extreme quantiles $q \in [0.90,1.00)$ and for sample sizes $N \in
\{2500,5000\}$. The solid curves indicate the true treatment effects.
Observe that the conventional estimator $\hat{\tau}_q^{CIC}$ tends to give
biased estimates as $q \rightarrow 1$. On the other hand, our proposed
estimator $\hat{\tau}_q^{eCIC}$ yields significantly less biased estimates
even in the limit $q \rightarrow 1$. We ran many additional simulations with
varying design parameter values $(\pi_G,\pi_T,\pi_A,\pi_B,\alpha)$, and the
results indicate similar patterns across sets of simulations.

\begin{figure}[t]
\centering
\begin{tabular}{cc}
\includegraphics[width=6.6cm]{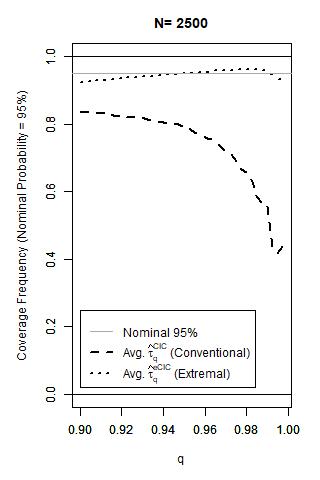}
& %
\includegraphics[width=6.6cm]{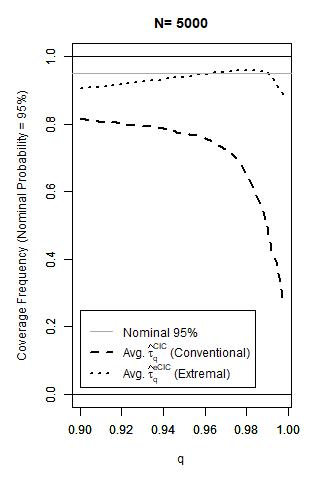}
\\ 
& 
\end{tabular}
${}$%
\caption{\setlength{\baselineskip}{5.5mm}Monte Carlo frequencies of coverage
of the true treatment effects by the 95\% confidence intervals at the
extreme quantiles $q \in [0.90,1.00)$ under the design with $(\protect\pi_G,%
\protect\pi_T,\protect\pi_A,\protect\pi_B,\protect\alpha%
)=(0.1,0.5,1.0,2.0,10)$. The dashed and dotted curves indicate the results
based on the conventional estimator $\hat{\protect\tau}_q^{CIC}$ and our
proposed estimator $\hat{\protect\tau}_q^{eCIC}$, respectively.}
\label{fig:sim:cover}
\end{figure}

Figure \ref{fig:sim:cover} shows Monte Carlo frequencies that the true
treatment effects are covered by the 95\% confidence intervals. The dashed
curves indicate the results based on the conventional estimator $\hat{\tau}%
_q^{CIC}$ and the dotted curves indicate the results based on our proposed
estimator $\hat{\tau}_q^{eCIC}$. The results are shown at the extreme
quantiles $q \in [0.90,1.00)$ and for sample sizes $N \in \{2500,5000\}$.
Observe that the coverage frequency based on the conventional method
deviates away from the nominal probability of 0.95 as $q \rightarrow 1$. In
contrast, the coverage frequency based on our proposed method is close to
the nominal probability of 0.95 at each point $q \in [0.90,1.00)$ in the
extreme quantiles. We remark again that we ran many additional simulations
with varying design parameter values $(\pi_G,\pi_T,\pi_A,\pi_B,\alpha)$, and
the results indicate similar patterns across sets of simulations.

In light of these simulation results, we provide the following practical recommendation. 
Use the conventional estimator $\hat{\tau}_q^{CIC}$ of \citet[][page 464]{athey2006identification} along with its standard error estimator \citep[][pages 464-465]{athey2006identification} for intermediate quantiles. 
On the other hand, use our proposed estimator $\hat{\tau}_q^{eCIC}$ in \eqref{eq:cic_est} along with the standard error estimator \eqref{eq:stderr} for extreme quantiles. 
The switching point can be chosen by using the log-log plot described in Section \ref{sec:switching}. 
We also follow this practical guideline for the empirical application to be presented in the next section.


\section{EITC and Extremely Low Birth Weights}

\label{sec:application} 
There is a long history in health economics research to study causes and
prevention of low infant birth weight. It is an important topic from policy
viewpoint because low infant birth weight has been identified to have
long-lasting impacts on the health and economic well being in adulthood %
\citep[e.g.,][]{currie2011inequality} as well as they are well known to have
immediate impact on infant mortality. Some economic and behavioral factors
affecting infant birth weight include maternal smoking %
\citep[e.g.,][]{almond2005costs,currie2009air}, maternal stress %
\citep[e.g.,][]{aizer2009maternal,camacho2008stress,evans2014giving}, and
economic resources \citep[e.g.,][]{hoynes2015income}, among others.

With studies of \textit{average} effects as in most of the existing
empirical studies, it still remains unknown if these causal factors would
have positive impacts on the most vulnerable subpopulation, namely those
infants born with extremely low birth weights. There are a few papers %
\citep{chernozhukov2011inference,sasaki2022fixed} that study extreme
quantiles of infant birth weights, but causal interpretations of their
estimation results require to assume exogeneity of the explanatory variable
of interest conditional on other observed covariates. In empirical settings
admitting a changes-in-changes design, on the other hand, we can handle
flexible endogeneity in the treatment choice and study treatment effects for
the most vulnerable subpopulation at the extremely low quantiles using the
method proposed in this paper.

\citet{hoynes2015income} use the difference-in-differences (DID) design
based on EITC reform (Omnibus Reconciliation Act of 1993, OBRA93) to
evaluate the effects of income gains through the EITC on infant health
outcomes. They find significant average effects of income shocks on the
incidence of low birth weight and the average infant birth weight. In this
paper, we aim to complement the work of \citet{hoynes2015income} by
analyzing the heterogeneous effects of the income gains through the EITC on infant birth
weight at extremely low quantiles, as opposed to those on average.

Following the prior work by \citet{hoynes2015income}, we use the U.S. Vital
Statistics Natality Data, 1989--1999. We also adopt their DID design for our
extreme CIC analysis by following their two key assumptions. First, the
effects of the EITC on infant birth weights run through the cash available
to the family which arrives through tax refunds and the cash is spent over
the subsequent 12 months. Second, we focus on the effects during the
sensitive development stage in the three months prior to birth.
Consequently, following the cash-in-hand assignment rule of \citet[][Table
1]{hoynes2015income}, we include births in May 1994 or after in the ``Post''
group ($T=1$) associated with the policy event of OBRA93. The eligibility
criteria for the EITC includes the requirement that a taxpayer has a
qualifying child. In this light, we include all the second- or higher-order
live births as the treatment group ($G=1$).
The sample sizes are  $n_{00}=2372001$, $n_{01}=1287185$, $n_{10}=2652321$, and $n_{11}=1325598$. 

\citet{hoynes2015income} define subpopulations by year, state, parity, education, race, and mother's age.
Then, they treat such a subpopulation as a unit of observation, and use the average birth weight within a subpopulation as the outcome value for the unit.
However, this procedure will not allow us to analyze individual heterogeneity with the quantile treatment effect because aggregation eliminates individual heterogeneity.
Hence, we use each birth as a unit of observation unlike \citet{hoynes2015income}.
Otherwise, we follow their empirical approach as follows.
First, we use year and state fixed effects.
Since \citet{hoynes2015income} use parity, education, race, and mother's age to define their subgroups of aggregation, we instead use this list of variables as covariates in our analysis.
To accommodate these covariates, the extended method introduced in Section \ref{sec:covariates} is employed.
Second, we focus on single women with high school education or less as in \citet{hoynes2015income}.

\begin{figure}[tbp]
\centering
\begin{tabular}{c}
\includegraphics[width=11.5cm]{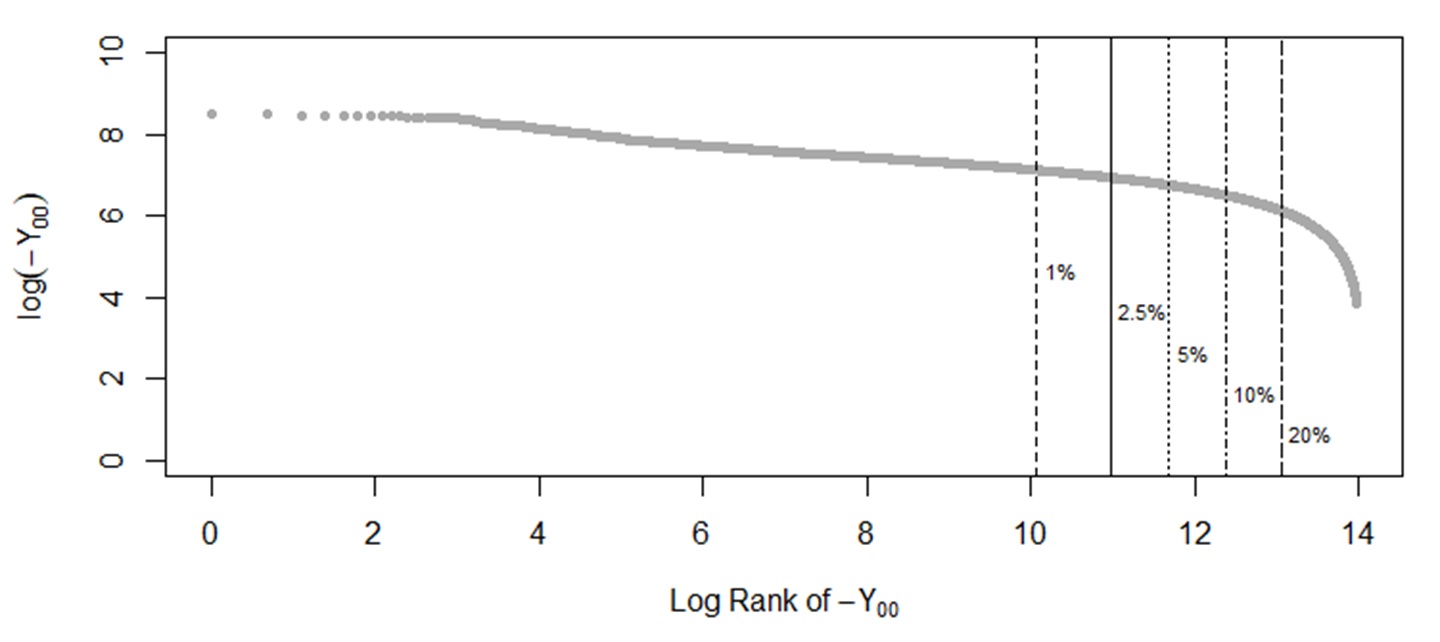}\\
\includegraphics[width=11.5cm]{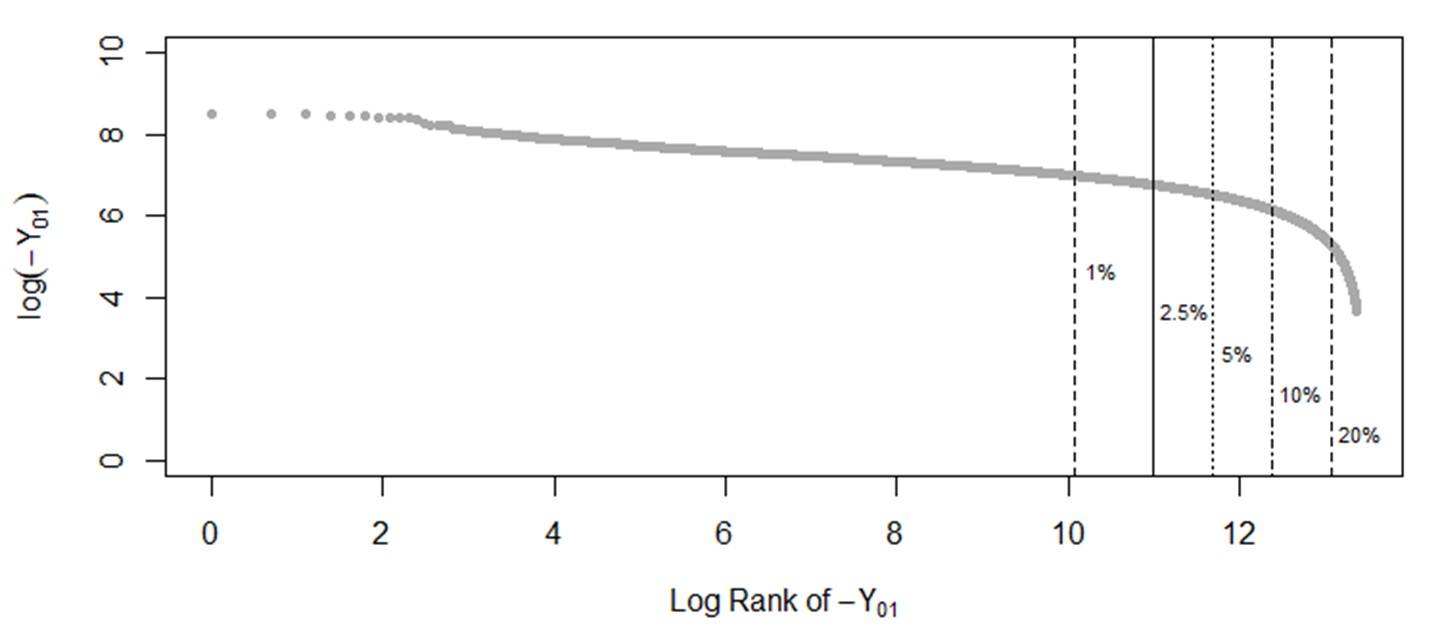}\\
\includegraphics[width=11.5cm]{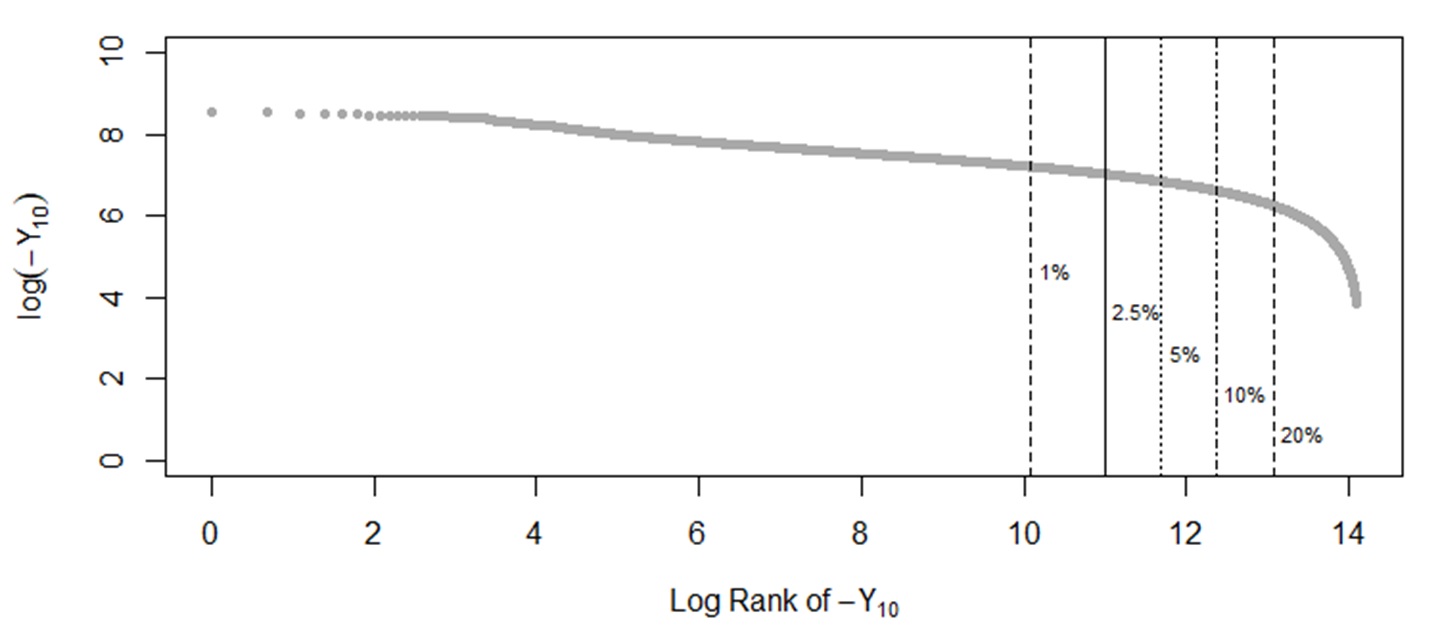}\\
\includegraphics[width=11.5cm]{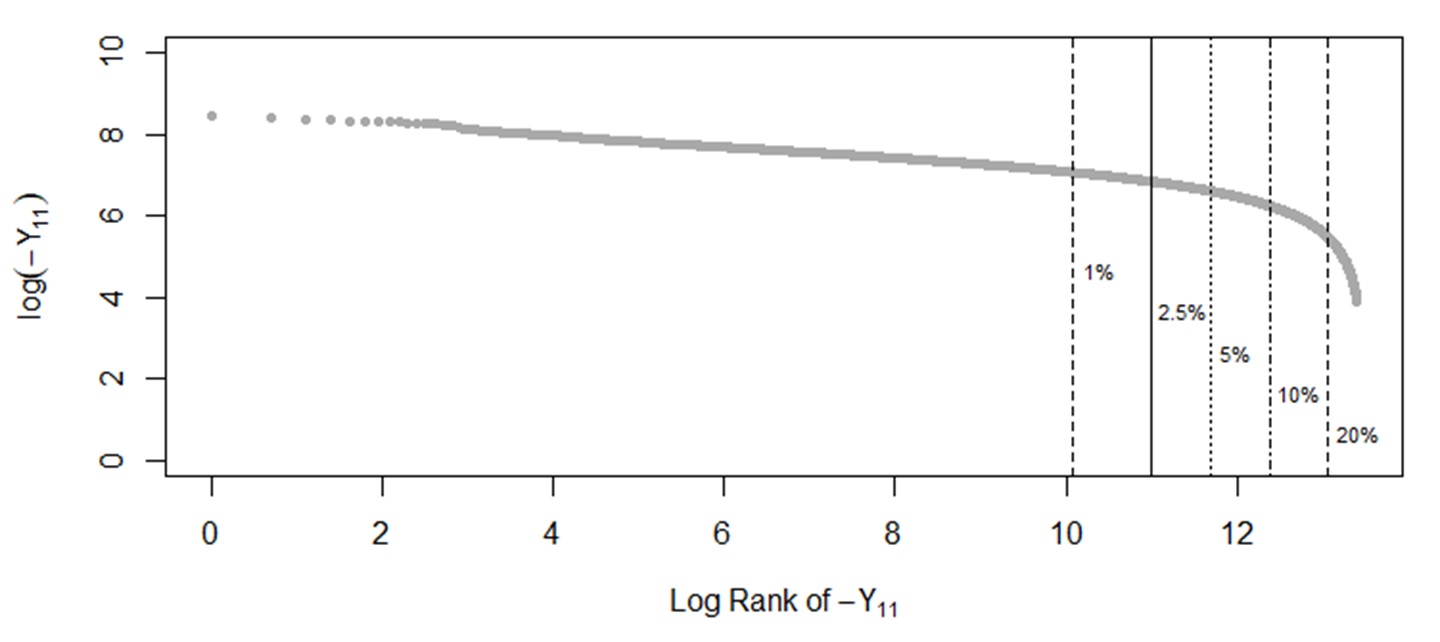}
\end{tabular}
${}$
\caption{\setlength{\baselineskip}{5.5mm} The log-log plots for $-\hat Y_{00}$, $-\hat Y_{01}$, $-\hat Y_{10}$, and $-\hat Y_{11}$.}
\label{fig:log_log}
\end{figure}

To determine the switching point $\overline q$ between our extreme CIC estimator and the conventional CIC estimator, we draw the log-log plots for $-\hat Y_{00}$, $-\hat Y_{01}$, $-\hat Y_{10}$, and $-\hat Y_{11}$ in Figure \ref{fig:log_log}.
Observe in each figure that the plot is reasonably linear up to around the 2.5-th or 5-th percentile, and thereby starts to curve downward.
In light of the discussion in Section \ref{sec:switching} and noting that our current focus is on the left tail, we choose the switching point $\overline q$ such that the log-log plot is linear for $i \in \{1,\cdots,\lfloor n_{gt} \overline q \rfloor\}$.
To guarantee a well Pareto tail approximation, we define the 2.5-th percentile as our switching point.

\begin{figure}[tbp]
\centering
\begin{tabular}{c}
\includegraphics[width=14.5cm]{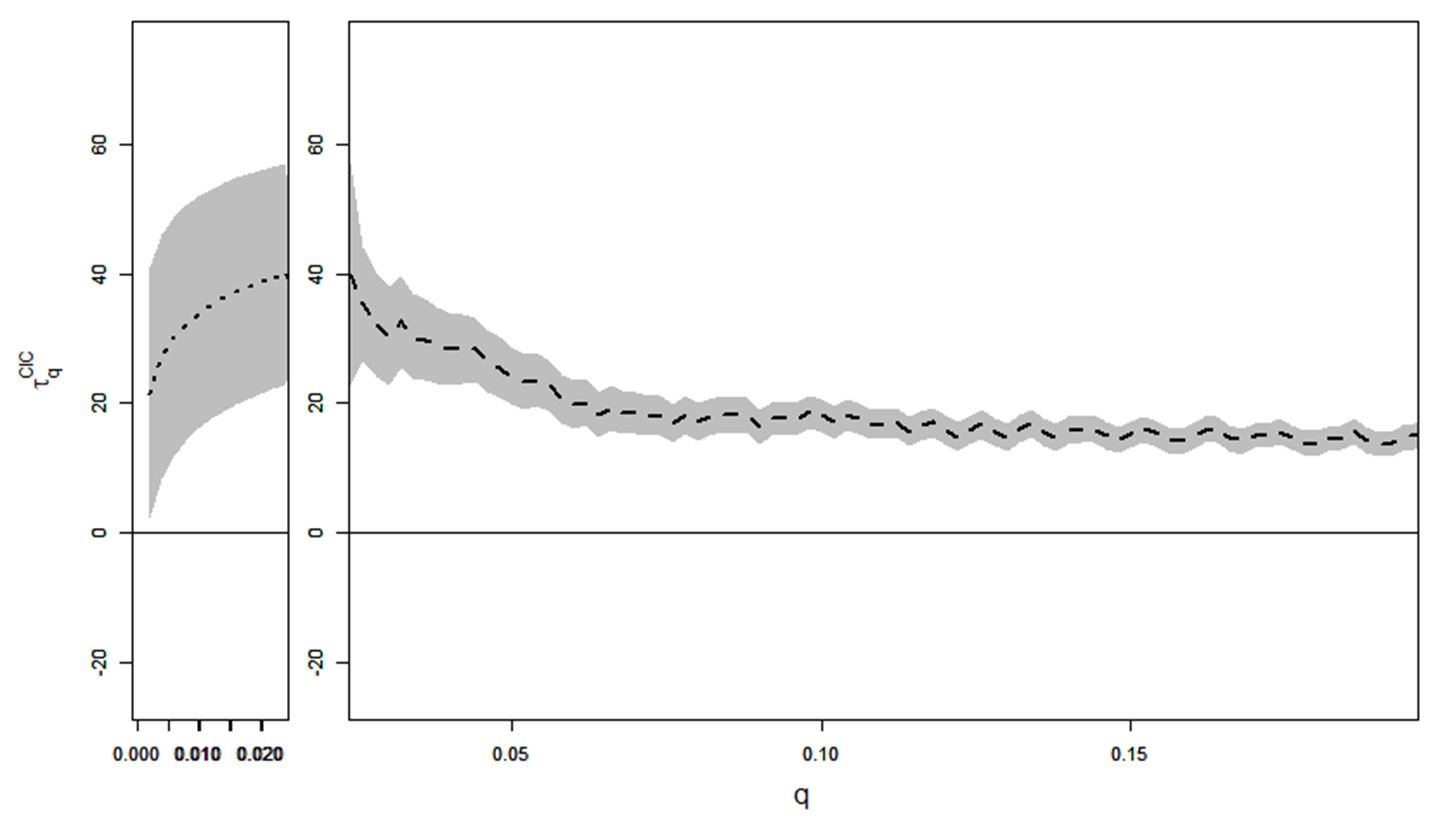}
\end{tabular}
${}$
\caption{\setlength{\baselineskip}{5.5mm}Estimates and 95 percent confidence intervals for $\protect\tau_q^{CIC}$ of infant birth weight for $q \in (0.000,0.200]$. 
The sample consists of infants born between 1989 and 1999 from unmarried black mothers who have complete 12 years of education. 
The results for the extreme quantiles $q \in (0.000, 0.025]$ are based on the proposed method. 
The results for the middle quantiles $q \in (0.025,0.200]$ are based on \citet{athey2006identification}.}
\label{fig:app}
\end{figure}

Figure \ref{fig:app} illustrates estimates and confidence intervals for $\tau_q^{CIC}$. 
The estimates by our proposed method for the extreme quantiles $q \in (0.000,0.025]$ are indicated by dotted curves, and the estimates by \citet{athey2006identification} for the intermediate quantiles $q \in (0.025,0.200]$ are indicated by the dashed curves. 
The gray shades indicate pointwise 95 percent confidence intervals.

Observe that the point estimates are unambiguously positive for all the quantiles $q \in (0.000,0.200)$. 
Furthermore, these income effects are statistically significant at each quantile $q \in (0.000,0.200]$. 
Therefore, we can conclude that income gains will causally improve the infant birth weights at low quantiles.

While \citet{hoynes2015income} discover positive effects of the EITC income gains \textit{on average}, we further find positive effects at the low quantiles in particular. 
This progress in empirical research is important as causal effects for extremely low infant birth weights are more relevant to policy analysis. 
Low infant birth weight is known to have have long-lasting impacts on the health and economic well being in adulthood \citep[e.g.,][]{currie2011inequality} as well as they are known to have immediate impact on infant mortality. 
Our findings focusing on the low quantiles imply that income support during pregnancy may help mitigate these adverse health and economic outcomes. 
We want to stress that, for us to reach this important empirical conclusion, both the conventional estimator $\hat\tau_q^{CIC}$ by \citet{athey2006identification} and our proposed estimator $\hat\tau_q^{eCIC}$ along with their standard errors are indispensable.

\section{Simulations Based on Empirical Data}\label{sec:simulations_empirical} 

Section \ref{sec:simulations} presents simulation studies based on data generated from an artificial design.
In this section, we present additional simulation studies with resamples from the empirical data which we use in Section \ref{sec:application}.

Let $\{\hat Y_{gt}^i\}_{i=1}^{n_{gt}}$ denote the residualized sample we obtain in Section \ref{sec:application} for each $g$ and $t$.
For each $g \in \{0,1\}$, we draw a one-percent subsample of size $\lfloor 0.01 \cdot n_{g0} \rfloor$ from $\{\hat Y_{00}^i\}_{i=1}^{n_{00}} \cup \{\hat Y_{10}^i\}_{i=1}^{n_{10}}$ with replacement, and define this subsample as a simulated sample of $Y_{g0}$.
Similarly, from each $g \in \{0,1\}$, we draw a one-percent subsample of size $\lfloor 0.01 \cdot n_{g1} \rfloor$ from $\{\hat Y_{01}^i\}_{i=1}^{n_{01}} \cup \{\hat Y_{11}^i\}_{i=1}^{n_{11}}$ with replacement, and define this subsample as a simulated sample of $Y_{g1}$.
Since we pool the source samples between the control and the treatment groups for each $t$, the true quantile treatment effect $\tau^{CIC}_q$ is zero for all $q$ by construction.
Recall from Section \ref{sec:application} that the original sample sizes are $n_{00}=2372001$, $n_{01}=1287185$, $n_{10}=2652321$, and $n_{11}=1325598$. 
Hence, simulation sample sizes are $\lfloor 0.01 \cdot n_{00} \rfloor=23720$, $\lfloor 0.01 \cdot n_{01} \rfloor=12871$, $\lfloor 0.01 \cdot n_{10} \rfloor=26523$, and $\lfloor 0.01 \cdot n_{11} \rfloor=13255$. 
Under this empirical Monte Carlo design, we run the same set of estimation and inference as in Section \ref{sec:simulations}, except that we focus on the left tail $q \in (0.00,0.10]$ as opposed to the right tail $q \in [0.90,1.00)$.

\begin{figure}[tbp]
\centering
\includegraphics[width=6.3cm]{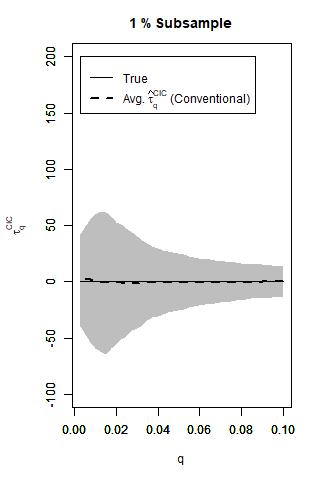}
\hspace{3cm}
\includegraphics[width=6.3cm]{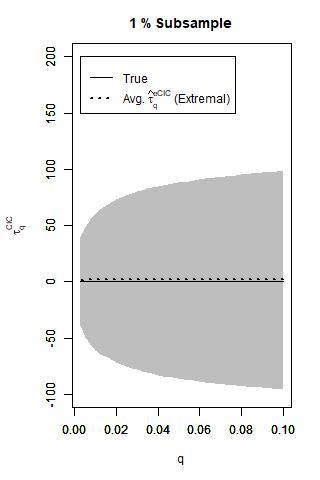}
\vspace{-0.5cm}\\
\includegraphics[width=6.3cm]{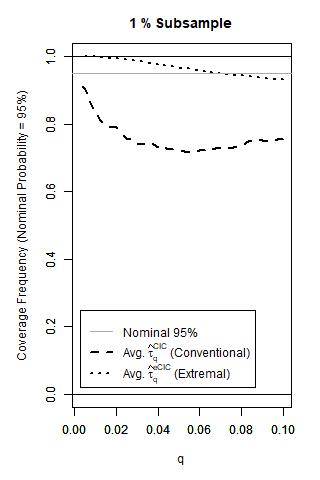}
\caption{\setlength{\baselineskip}{5.5mm}Top: Monte Carlo averages and inter-quartile ranges (shaded) of the estimates based on the conventional estimator $\hat{\protect\tau}_q^{CIC}$ (dashed curves on the left column)
and our proposed estimator $\hat{\protect\tau}_q^{eCIC}$ (dotted curves on the right column) at the extreme quantiles $q \in (0.00, 0.10]$. The true treatment effects are indicated by the solid curves.
Bottom: Monte Carlo frequencies of coverage
of the true treatment effects by the 95\% confidence intervals at the extreme quantiles $q \in (0.00,0.10]$. The dashed and dotted curves indicate the results based on the conventional estimator $\hat{\protect\tau}_q^{CIC}$ and our proposed estimator $\hat{\protect\tau}_q^{eCIC}$, respectively.}
\label{fig:sim_empirical}
\end{figure}

The top row of Figure \ref{fig:sim_empirical} shows Monte Carlo averages and inter-quartile ranges of the estimates, analogously to Figure \ref{fig:sim:estimates} in Section \ref{sec:simulations}.
The dashed curve on the left panel indicates the average estimates based on the conventional estimator $\hat{\tau}_q^{CIC}$. 
The dotted curve on the right panel indicates the average estimates based on our proposed estimator $\hat{\tau}_q^{eCIC}$. 
In each panel, the shaded region indicates the inter-quartile ranges of the estimates by the respective methods. 
The solid curves indicate the true treatment effects.
Since the true treatment effects are homogeneously zero for all $q$ under the current data generating design, there is little bias in the both estimators.
Therefore, the inter-quartile ranges are nicely symmetric for the both estimators.
This feature of the results contrasts with that in Section \ref{sec:simulations}, where non-trivial biases exist for the conventional estimator $\hat\tau_q^{CIC}$ at the extreme quantiles.

The bottom row of Figure \ref{fig:sim_empirical} shows Monte Carlo frequencies that the true treatment effects are covered by the 95\% confidence intervals, analogously to Figure \ref{fig:sim:cover} in Section \ref{sec:simulations}.
The dashed curve indicates the results based on the conventional estimator $\hat{\tau}_q^{CIC}$ and the dotted curve indicates the results based on our proposed estimator $\hat{\tau}_q^{eCIC}$. 
The results are shown at the extreme quantiles $q \in (0.00,0.10]$.
Although the conventional estimator does not suffer from bias under the current design, its statistical inference still suffers from size distortions.
Our proposed extreme CIC estimator $\hat\tau_q^{eCIC}$ yields substantially less size distortions than the conventional estimator $\hat\tau_q^{CIC}$.

\section{Summary and Discussions}\label{sec:conclusion} 

In this paper, we propose a new CIC estimator to accurately estimate the
treatment effects at extreme/tail quantiles. We also derive its asymptotic
normality result for statistical inference. Our proposal of these new
methods is motivated by the fact that policy analysts are often interested
in treating subpopulations near tails of the distributions of outcome
variables (e.g., extremely poor individuals and infants with extremely low
birth weights) while existing CIC estimators are tailored to middle
quantiles.

Simulation studies demonstrate that the new extreme CIC estimator along with
its standard error estimator performs better than the conventional method in
the tails. Based on our observations of these results, we propose to use our
proposed CIC estimator for extreme quantiles, while the conventional CIC
estimation should be used for intermediate quantiles.

Applying the proposed method to U.S. Vital Statistics Natality Data, we
study the effects of income gains from the 1993 EITC reform on infant birth
weights for those in the most critical conditions. We find significant
positive effects of the income gains on infant birth weights for the subpopulation 
at the low quantiles of birth weight. 


Finally, we remind the readers that this paper is accompanied by a Stata command, \texttt{ecic} (extreme changes in changes).
The package can be installed from SSC archive with the following command line: \underline{\texttt{ssc} \texttt{install} \texttt{ecic}}.
After the installation, run \underline{\texttt{help} \texttt{ecic}} for usage of the command.

\vspace{0cm} 
\bibliographystyle{ecta}
\bibliography{bib}

\begin{thebibliography}{26}
\newcommand{\enquote}[1]{``#1''}
\expandafter\ifx\csname natexlab\endcsname\relax\def\natexlab#1{#1}\fi

\bibitem[\protect\citeauthoryear{Aizer, Stroud, and Buka}{Aizer
  et~al.}{2009}]{aizer2009maternal}
\textsc{Aizer, A., L.~Stroud, and S.~Buka} (2009): \enquote{Maternal stress and
  child well-being: Evidence from siblings,} Unpublished Manuscript, Brown
  University, Providence, RI.

\bibitem[\protect\citeauthoryear{Almond, Chay, and Lee}{Almond
  et~al.}{2005}]{almond2005costs}
\textsc{Almond, D., K.~Y. Chay, and D.~S. Lee} (2005): \enquote{The costs of
  low birth weight,} \emph{Quarterly Journal of Economics}, 120, 1031--1083.

\bibitem[\protect\citeauthoryear{Athey and Imbens}{Athey and
  Imbens}{2006}]{athey2006identification}
\textsc{Athey, S. and G.~W. Imbens} (2006): \enquote{Identification and
  inference in nonlinear difference-in-differences models,}
  \emph{Econometrica}, 74, 431--497.

\bibitem[\protect\citeauthoryear{Camacho}{Camacho}{2008}]{camacho2008stress}
\textsc{Camacho, A.} (2008): \enquote{Stress and birth weight: evidence from
  terrorist attacks,} \emph{American Economic Review}, 98, 511--15.

\bibitem[\protect\citeauthoryear{Carpentier and Kim}{Carpentier and
  Kim}{2014}]{CarpentierKim2014}
\textsc{Carpentier, A. and A.~K.~H. Kim} (2014): \enquote{Adaptive and minimax
  optimal estimation of the tail coefficient,} \emph{Statistica Sinica}, 25,
  1133--1144.

\bibitem[\protect\citeauthoryear{Cheng and Peng}{Cheng and
  Peng}{2001}]{cheng2001confidence}
\textsc{Cheng, S. and L.~Peng} (2001): \enquote{Confidence intervals for the
  tail index,} \emph{Bernoulli}, 7, 751--760.

\bibitem[\protect\citeauthoryear{Chernozhukov}{Chernozhukov}{2005}]{chernozhukov2005extremal}
\textsc{Chernozhukov, V.} (2005): \enquote{Extremal quantile regression,}
  \emph{Annals of Statistics}, 806--839.

\bibitem[\protect\citeauthoryear{Chernozhukov and
  Fern{\'a}ndez-Val}{Chernozhukov and
  Fern{\'a}ndez-Val}{2011}]{chernozhukov2011inference}
\textsc{Chernozhukov, V. and I.~Fern{\'a}ndez-Val} (2011): \enquote{Inference
  for extremal conditional quantile models, with an application to market and
  birthweight risks,} \emph{Review of Economic Studies}, 78, 559--589.

\bibitem[\protect\citeauthoryear{Currie}{Currie}{2011}]{currie2011inequality}
\textsc{Currie, J.} (2011): \enquote{Inequality at birth: some causes and
  consequences,} \emph{American Economic Review}, 101, 1--22.

\bibitem[\protect\citeauthoryear{Currie, Neidell, and Schmieder}{Currie
  et~al.}{2009}]{currie2009air}
\textsc{Currie, J., M.~Neidell, and J.~F. Schmieder} (2009): \enquote{Air
  pollution and infant health: Lessons from New Jersey,} \emph{Journal of
  Health Economics}, 28, 688--703.

\bibitem[\protect\citeauthoryear{de~Chaisemartin and
  D'Haultf{\oe}uille}{de~Chaisemartin and
  D'Haultf{\oe}uille}{2014}]{de2014fuzzy}
\textsc{de~Chaisemartin, C. and X.~D'Haultf{\oe}uille} (2014): \enquote{Fuzzy
  changes-in-changes,} Unpublished Manuscript.

\bibitem[\protect\citeauthoryear{de~Haan and Ferreira}{de~Haan and
  Ferreira}{2007}]{de2007extreme}
\textsc{de~Haan, L. and A.~Ferreira} (2007): \emph{Extreme Value Theory: An
  Introduction}, Springer Science \& Business Media.

\bibitem[\protect\citeauthoryear{Deuber, Li, Engelke, and Maathuis}{Deuber
  et~al.}{2021}]{deuber2021estimation}
\textsc{Deuber, D., J.~Li, S.~Engelke, and M.~H. Maathuis} (2021):
  \enquote{Estimation and inference of extremal quantile treatment effects for
  heavy-tailed distributions,} \emph{arXiv preprint arXiv:2110.06627}.

\bibitem[\protect\citeauthoryear{D'Haultf{\oe}uille, Hoderlein, and
  Sasaki}{D'Haultf{\oe}uille et~al.}{2022}]{dhaultoeuille2022nonparametric}
\textsc{D'Haultf{\oe}uille, X., S.~Hoderlein, and Y.~Sasaki} (2022):
  \enquote{Nonparametric difference-in-differences in repeated cross-sections
  with continuous treatments,} \emph{Journal of Econometrics}, forthcoming.

\bibitem[\protect\citeauthoryear{D'Haultf{\oe}uille, Maurel, and
  Zhang}{D'Haultf{\oe}uille et~al.}{2018}]{d2018extremal}
\textsc{D'Haultf{\oe}uille, X., A.~Maurel, and Y.~Zhang} (2018):
  \enquote{Extremal quantile regressions for selection models and the
  black--white wage gap,} \emph{Journal of Econometrics}, 203, 129--142.

\bibitem[\protect\citeauthoryear{Evans and Garthwaite}{Evans and
  Garthwaite}{2014}]{evans2014giving}
\textsc{Evans, W.~N. and C.~L. Garthwaite} (2014): \enquote{Giving mom a break:
  The impact of higher EITC payments on maternal health,} \emph{American
  Economic Journal: Economic Policy}, 6, 258--90.

\bibitem[\protect\citeauthoryear{Ghanem, Hirshleifer, Kedagni, and
  Ortiz-Becerra}{Ghanem et~al.}{2022}]{ghanem2022correcting}
\textsc{Ghanem, D., S.~Hirshleifer, D.~Kedagni, and K.~Ortiz-Becerra} (2022):
  \enquote{Correcting Attrition Bias using Changes-in-Changes,} \emph{arXiv
  preprint arXiv:2203.12740}.

\bibitem[\protect\citeauthoryear{Girard, Stupfler, and Usseglio-Carleve}{Girard
  et~al.}{2021}]{girard2021}
\textsc{Girard, S., G.~Stupfler, and A.~Usseglio-Carleve} (2021):
  \enquote{Extreme conditional expectile estimation in heavy-tailed
  heteroscedastic regression models,} \emph{Annals of Statistics}, 49,
  3358--3382.

\bibitem[\protect\citeauthoryear{Guillou and Hall}{Guillou and
  Hall}{2001}]{guillou2001diagnostic}
\textsc{Guillou, A. and P.~Hall} (2001): \enquote{A diagnostic for selecting
  the threshold in extreme value analysis,} \emph{Journal of the Royal
  Statistical Society: Series B (Statistical Methodology)}, 63, 293--305.

\bibitem[\protect\citeauthoryear{Haeusler and Segers}{Haeusler and
  Segers}{2007}]{haeusler2007}
\textsc{Haeusler, E. and J.~Segers} (2007): \enquote{Assessing confidence
  intervals for the tail index by Edgeworth expansions for the Hill estimator,}
  \emph{Bernoulli}, 13, 175--194.

\bibitem[\protect\citeauthoryear{Hill}{Hill}{1975}]{hill1975simple}
\textsc{Hill, B.~M.} (1975): \enquote{A simple general approach to inference
  about the tail of a distribution,} \emph{Annals of Statistics}, 1163--1174.

\bibitem[\protect\citeauthoryear{Hoynes, Miller, and Simon}{Hoynes
  et~al.}{2015}]{hoynes2015income}
\textsc{Hoynes, H., D.~Miller, and D.~Simon} (2015): \enquote{Income, the
  earned income tax credit, and infant health,} \emph{American Economic
  Journal: Economic Policy}, 7, 172--211.

\bibitem[\protect\citeauthoryear{Melly and Santangelo}{Melly and
  Santangelo}{2015}]{melly2015changes}
\textsc{Melly, B. and G.~Santangelo} (2015): \enquote{The changes-in-changes
  model with covariates,} Unpublished Manuscript, Universit{\"a}t Bern, Bern.

\bibitem[\protect\citeauthoryear{Resnick}{Resnick}{2007}]{resnick2007book}
\textsc{Resnick, S.~I.} (2007): \emph{Heavy-tail phenomena: probabilistic and
  statistical modeling}, Springer Science \& Business Media.

\bibitem[\protect\citeauthoryear{Sasaki and Wang}{Sasaki and
  Wang}{2022}]{sasaki2022fixed}
\textsc{Sasaki, Y. and Y.~Wang} (2022): \enquote{Fixed-k inference for
  conditional extremal quantiles,} \emph{Journal of Business \& Economic
  Statistics}, 40, 829--837.

\bibitem[\protect\citeauthoryear{Zhang}{Zhang}{2018}]{zhang2018extremal}
\textsc{Zhang, Y.} (2018): \enquote{Extremal quantile treatment effects,}
  \emph{Annals of Statistics}, 46, 3707--3740.

\end{thebibliography}
\vspace{0cm}


\section*{Appendix}

\appendix

\section{Proof of Theorem \protect\ref{thm:cic}}

\label{sec:thm:cic} 

\begin{proof}
For succinctness, we use the short-hand notation $F_{gt}\left( \cdot \right) 
$ for $F_{Y_{gt}}\left( \cdot \right) $, and accordingly use the short-hand
notation $F_{gt}^{-1}\left( \cdot \right) $ for $F_{Y_{gt}}^{-1}\left( \cdot
\right) $. Under Conditions 1, 2, and 4, we have 
\begin{equation}
\sqrt{k_{gt}}\left( \hat{\alpha}_{gt}-\alpha _{gt}\right) \equiv \Gamma _{gt}%
\overset{d}{\rightarrow }\mathcal{N}\left( 0,\alpha _{gt}^{2}\right)
\label{asym:index}
\end{equation}%
for all $\left( g,t\right) \in \{0,1\}^{2}$ -- see \citet{hill1975simple}.
Moreover, under Conditions 1, 2, 4, and 5, we have 
\begin{equation}
\frac{\sqrt{k_{gt}}}{\log d_{gt}}\left( \frac{\hat{F}_{gt}^{-1}\left(
q\right) }{F_{gt}^{-1}\left( q\right) }-1\right) \equiv \Lambda _{gt}\overset%
{d}{\rightarrow }\mathcal{N}\left( 0,\alpha _{gt}^{-2}\right) ,
\label{asym:quantile}
\end{equation}%
for all $\left( g,t\right) \in \{0,1\}^{2}$ by Theorem 4.3.8 in %
\citet{de2007extreme}, where $d_{gt}\equiv k_{gt}/\left( n_{gt}\left(
1-q\right) \right) $. Given the indepence of $\{Y_{gt}\}$ across $g$ and $t$
under Condition 1, $\{\hat{F}_{Y_{gt}}^{-1}\left( q\right) ,\hat{\alpha}%
_{gt}\}$ are also independent across $\{g,t\}$. Thus, it suffices to derive
the limit of the second item in (\ref{eq:cic_est}), that is, 
\begin{eqnarray*}
\hat{A}_{n} &\equiv &\hat{F}_{01}^{-1}\left( \hat{F}_{00}\left( \hat{F}%
_{10}^{-1}\left( q\right) \right) \right) \\
&=&Y_{01}^{\left( k_{01}+1\right) }\left( \frac{Y_{10}^{\left(
k_{10}+1\right) }}{Y_{00}^{\left( k_{00}+1\right) }}\right) ^{\hat{\alpha}%
_{00}/\hat{\alpha}_{01}}\left( \frac{k_{01}}{n_{01}}\frac{n_{00}}{k_{00}}%
\right) ^{1/\hat{\alpha}_{01}}\left( \frac{k_{10}}{n_{10}}\right) ^{\hat{%
\alpha}_{00}/\left( \hat{\alpha}_{10}\hat{\alpha}_{01}\right) }\left(
1-q\right) ^{-\hat{\alpha}_{00}/\left( \hat{\alpha}_{10}\hat{\alpha}%
_{01}\right) }
\end{eqnarray*}

We also write the population counterpart as 
\begin{equation*}
A_{n}=F_{01}^{-1}\left( 1-\frac{k_{01}}{n_{01}}\right) \left( \frac{%
F_{10}^{-1}\left( 1-\frac{k_{10}}{n_{10}}\right) }{F_{00}^{-1}\left( 1-\frac{%
k_{00}}{n_{00}}\right) }\right) ^{\alpha _{00}/\alpha _{01}}\left( \frac{%
k_{01}}{n_{01}}\frac{n_{00}}{k_{00}}\right) ^{1/\alpha _{01}}\left( \frac{%
k_{10}}{n_{10}}\right) ^{\alpha _{00}/\left( \alpha _{10}\alpha _{01}\right)
}\left( 1-q\right) ^{-\alpha _{00}/\left( \alpha _{10}\alpha _{01}\right) },
\end{equation*}%
and we are going to linearize $\hat{A}_{n}/A_{n}-1$ around zero. First, note
that we have 
\begin{equation}
\sqrt{k_{gt}}\left( \frac{Y_{gt}^{\left( k_{gt}+1\right) }}{%
F_{gt}^{-1}\left( 1-k_{gt}/n_{gt}\right) }-1\right) \equiv \Delta _{gt}%
\overset{d}{\rightarrow }\mathcal{N}\left( 0,\alpha _{gt}^{-2}\right)
\label{asym:emp}
\end{equation}%
from Theorem 2.4.8 in \citet{de2007extreme} and our Condition 4. Second, we
decompose $\hat{A}_{n}/A_{n}$ as 
\begin{eqnarray*}
\log \left( \hat{A}_{n}/A_{n}\right) &=&\log \left( \frac{Y_{01}^{\left(
k_{01}+1\right) }}{F_{01}^{-1}\left( 1-\frac{k_{01}}{n_{01}}\right) }\right)
\\
&&+\left[ \frac{\hat{\alpha}_{00}}{\hat{\alpha}_{01}}\log \left( \frac{%
Y_{10}^{\left( k_{10}+1\right) }}{Y_{00}^{\left( k_{00}+1\right) }}\right) -%
\frac{\alpha _{00}}{\alpha _{01}}\log \left( \frac{F_{10}^{-1}\left( 1-\frac{%
k_{10}}{n_{10}}\right) }{F_{00}^{-1}\left( 1-\frac{k_{00}}{n_{00}}\right) }%
\right) \right] \\
&&+\left( \frac{1}{\hat{\alpha}_{01}}-\frac{1}{\alpha _{01}}\right) \log
\left( \frac{k_{01}}{n_{01}}\frac{n_{00}}{k_{00}}\right) \\
&&+\left( \frac{\hat{\alpha}_{00}}{\hat{\alpha}_{10}\hat{\alpha}_{01}}-\frac{%
\alpha _{00}}{\alpha _{10}\alpha _{01}}\right) \log \left( \frac{k_{10}}{%
n_{10}\left( 1-q\right) }\right) \\
&\equiv &I_{1n}+I_{2n}+I_{3n}+I_{4n}\text{.}
\end{eqnarray*}%
For the first term, $I_{1n}$, we have 
\begin{equation*}
I_{1n}=\log \left( \frac{Y_{01}^{\left( k_{01}+1\right) }}{F_{01}^{-1}\left(
1-\frac{k_{01}}{n_{01}}\right) }\right) =\log \left( 1+k_{01}^{-1/2}\Delta
_{01}\right) =k_{01}^{-1/2}\Delta _{01}+o_{p}\left( k_{01}^{-1/2}\right)
\end{equation*}%
by (\ref{asym:emp}). For the second term, $I_{2n}$, we decompose it as 
\begin{eqnarray*}
I_{2n} &=&\left( \frac{\hat{\alpha}_{00}}{\hat{\alpha}_{01}}-\frac{\alpha
_{00}}{\alpha _{01}}\right) \log \left( \frac{F_{10}^{-1}\left( 1-\frac{%
k_{10}}{n_{10}}\right) }{F_{00}^{-1}\left( 1-\frac{k_{00}}{n_{00}}\right) }%
\right) \\
&&+\frac{\hat{\alpha}_{00}}{\hat{\alpha}_{01}}\left[ \log \left( \frac{%
Y_{10}^{\left( k_{10}+1\right) }}{Y_{00}^{\left( k_{00}+1\right) }}\right)
-\log \left( \frac{F_{10}^{-1}\left( 1-\frac{k_{10}}{n_{10}}\right) }{%
F_{00}^{-1}\left( 1-\frac{k_{00}}{n_{00}}\right) }\right) \right] \\
&=&\left[ \frac{\left( \hat{\alpha}_{00}-\alpha _{00}\right) }{\hat{\alpha}%
_{01}}-\frac{\alpha _{00}}{\hat{\alpha}_{01}\alpha _{01}}\left( \hat{\alpha}%
_{01}-\alpha _{01}\right) \right] \log \left( \frac{F_{10}^{-1}\left( 1-%
\frac{k_{10}}{n_{10}}\right) }{F_{00}^{-1}\left( 1-\frac{k_{00}}{n_{00}}%
\right) }\right) \\
&&+\frac{\hat{\alpha}_{00}}{\hat{\alpha}_{01}}\left[ \log \left( \frac{%
Y_{10}^{\left( k_{10}+1\right) }}{F_{10}^{-1}\left( 1-\frac{k_{10}}{n_{10}}%
\right) }\right) -\log \left( \frac{Y_{00}^{\left( k_{00}+1\right) }}{%
F_{00}^{-1}\left( 1-\frac{k_{00}}{n_{00}}\right) }\right) \right] \\
&=&\left[ \frac{k_{00}^{-1/2}\Gamma _{00}}{\alpha _{01}+O_{p}\left(
k_{01}^{-1/2}\right) }-\frac{\alpha _{00}}{\alpha _{01}^{2}+O_{p}\left(
k_{01}^{-1/2}\right) }k_{01}^{-1/2}\Gamma _{01}\right] \log \left( \frac{%
F_{10}^{-1}\left( 1-\frac{k_{10}}{n_{10}}\right) }{F_{00}^{-1}\left( 1-\frac{%
k_{00}}{n_{00}}\right) }\right) \\
&&+\left( \frac{\alpha _{00}}{\alpha _{01}}+O_{p}\left(
k_{00}^{-1/2}+k_{01}^{-1/2}\right) \right) \left[ k_{10}^{-1/2}\Delta
_{10}-k_{00}^{-1/2}\Delta _{00}+o_{p}\left(
k_{10}^{-1/2}+k_{00}^{-1/2}\right) \right] .
\end{eqnarray*}%
by (\ref{asym:index}) and (\ref{asym:emp}). For term $I_{3n}$, we rewrite it
as 
\begin{equation*}
I_{3n}=\left[ k_{01}^{-1/2}\Gamma _{01}+o_{p}\left( k_{01}^{-1/2}\right) %
\right] \log \left( \frac{k_{01}}{n_{01}}\frac{n_{00}}{k_{00}}\right)
\end{equation*}%
by (\ref{asym:index}). For term $I_{4n}$, we rewrite it as 
\begin{equation*}
I_{4n}=\left( 
\begin{array}{c}
\frac{k_{00}^{-1/2}\Gamma _{00}}{\alpha _{10}\alpha _{01}}-\frac{\alpha _{00}%
}{\alpha _{10}^{2}\alpha _{01}}k_{10}^{-1/2}\Gamma _{10} \\ 
-\frac{\alpha _{00}}{\alpha _{10}\alpha _{01}^{2}}k_{01}^{-1/2}\Gamma
_{01}+o_{p}\left( k_{00}^{-1/2}+k_{10}^{-1/2}+k_{01}^{-1/2}\right)%
\end{array}%
\right) \log \left( \frac{k_{10}}{n_{10}\left( 1-q\right) }\right)
\end{equation*}%
by (\ref{asym:index}).

Conditions 4 and 5 imply that $d_{gt}=k_{gt}/\left[ n_{gt}\left(
1-q\right) \right] \rightarrow \infty $ for all $g,t\in \{0,1\}^{2}$.
Moreover, Condition 2 implies that $F_{gt}^{-1}\left( 1-\frac{k_{gt}}{%
n_{gt}}\right) = O(\left( k_{gt}/n_{gt}\right) ^{-1/\alpha _{gt}})$ for
all $g,t$. Then using L'Hospital's rule, Condition 3, and $d_{gt}
\rightarrow \infty $, we obtain that%
\begin{eqnarray*}
&&\frac{1}{\log \left( d_{11}\right) }\log \left( \frac{F_{10}^{-1}\left( 1-%
\frac{k_{10}}{n_{10}}\right) }{F_{00}^{-1}\left( 1-\frac{k_{00}}{n_{00}}%
\right) }\right) \\
&= &O\left( -\frac{\alpha _{10}^{-1}\log \left( k_{10}/n_{10}\right) }{\log
\left( k_{11}/n_{11}\right) -\log \left( 1-q\right) }+\frac{\alpha
_{00}^{-1}\log \left( k_{00}/n_{00}\right) }{\log \left(
k_{11}/n_{11}\right) -\log \left( 1-q\right) } \right) \\
&=&o(1).
\end{eqnarray*}

Now using the above derivations, 
we obtain 
\begin{eqnarray*}
\frac{\sqrt{k_{11}}}{\log \left( d_{11}\right) }I_{1n} &=&O_{p}\left( \frac{1%
}{\log \left( d_{11}\right) }\right) =o_{p}(1) \\
\frac{\sqrt{k_{11}}}{\log \left( d_{11}\right) }I_{2n} &=&O_{p}\left( \frac{1%
}{\log \left( d_{11}\right) }\log \left( \frac{F_{10}^{-1}\left( 1-\frac{%
k_{10}}{n_{10}}\right) }{F_{00}^{-1}\left( 1-\frac{k_{00}}{n_{00}}\right) }%
\right) \right) =o_{p}(1) \\
\frac{\sqrt{k_{11}}}{\log \left( d_{11}\right) }I_{3n} &=&O_{p}\left( \frac{1%
}{\log \left( d_{11}\right) }\frac{k_{01}}{n_{01}}\frac{n_{00}}{k_{00}}%
\right) =o_{p}(1) \\
\frac{\sqrt{k_{11}}}{\log \left( d_{11}\right) }I_{4n} &=&\frac{\log \left(
d_{10}\right) }{\log \left( d_{11}\right) }\left[ \left( \frac{k_{11}}{k_{00}%
}\right) ^{1/2}\frac{\Gamma _{00}}{\alpha _{10}\alpha _{01}}-\left( \frac{%
k_{11}}{k_{10}}\right) ^{1/2}\frac{\alpha _{00}\Gamma _{10}}{\alpha
_{10}^{2}\alpha _{01}}-\left( \frac{k_{11}}{k_{01}}\right) ^{1/2}\frac{%
\alpha _{00}\Gamma _{01}}{\alpha _{10}\alpha _{01}^{2}}\right] +o_{p}\left(
1\right) .
\end{eqnarray*}%
Now, combining $I_{1n}$, $I_{2n}$, $I_{3n}$, and $I_{4n}$, and using the
fact that $\exp (x)=1+x+O(x^{2})$ as $x\rightarrow 0$, we obtain 
\begin{eqnarray*}
&&\frac{\sqrt{k_{11}}}{\log \left( d_{11}\right) }\left( \frac{\hat{A}_{n}}{%
A_{n}}-1\right) \\
&=&\frac{\log \left( d_{10}\right) }{\log \left( d_{11}\right) }\left[
\left( \frac{k_{11}}{k_{00}}\right) ^{1/2}\frac{\Gamma _{00}}{\alpha
_{10}\alpha _{01}}-\left( \frac{k_{11}}{k_{10}}\right) ^{1/2}\frac{\alpha
_{00}\Gamma _{10}}{\alpha _{10}^{2}\alpha _{01}}-\left( \frac{k_{11}}{k_{01}}%
\right) ^{1/2}\frac{\alpha _{00}\Gamma _{01}}{\alpha _{10}\alpha _{01}^{2}}%
\right] +o_{p}\left( 1\right) \\
&&\overset{d}{\rightarrow }\frac{\lambda _{11/10}}{\eta _{11/10}}\mathcal{N}%
\left( 0,\lambda _{11/00}\frac{\alpha _{00}^{2}}{\alpha _{10}^{2}\alpha
_{01}^{2}}+\lambda _{11/10}\frac{\alpha _{00}^{2}}{\alpha _{01}^{2}\alpha
_{01}^{2}}+\lambda _{11/01}\frac{\alpha _{00}^{2}}{\alpha _{10}^{2}\alpha
_{01}^{2}}\right) .
\end{eqnarray*}%
by independence among $\Gamma _{00}$, $\Gamma _{10}$ and $\Gamma _{01}$,
Condition 3, and \eqref{asym:index}. 

Finally, using (\ref{asym:quantile}) with $g=t=1$ and the condition that $%
F_{11}^{-1}\left( q\right) /A_{n}\rightarrow \varsigma $, we obtain 
\begin{eqnarray*}
\frac{k_{11}^{1/2}}{\log d_{11}}\left( \frac{\hat{\tau}_{q}^{eCIC}-\tau
_{q}^{eCIC}}{F_{11}^{-1}\left( q\right) }\right)  &=&\frac{k_{11}^{1/2}}{%
\log d_{11}}\left( \frac{\hat{F}_{11}^{-1}\left( q\right) -F_{11}^{-1}\left(
q\right) }{F_{11}^{-1}\left( q\right) }-\frac{\hat{A}_{n}-A_{n}}{%
F_{11}^{-1}\left( q\right) }\right)  \\
&=&\frac{k_{11}^{1/2}}{\log d_{11}}\left( \frac{\hat{F}_{11}^{-1}\left(
q\right) }{F_{11}^{-1}\left( q\right) }-1\right)  \\
&&-\frac{k_{11}^{1/2}}{\log d_{11}}\left( \frac{\hat{A}_{n}}{A_{n}}-1\right)
\left( \frac{A_{n}}{F_{11}^{-1}\left( q\right) }\right)  \\
&\overset{d}{\rightarrow }&\mathcal{N}\left( 0,\Omega \right) ,
\end{eqnarray*}%
where%
\begin{equation*}
\Omega =\alpha _{11}^{-2}+\left( \frac{1}{\varsigma }\right) ^{2}\left( 
\frac{\lambda _{11/10}}{\eta _{11/10}}\right) ^{2}\left[ \lambda
_{11/00}+\lambda _{11/10}+\lambda _{11/01}\right] \frac{\alpha _{00}^{2}}{%
\alpha _{10}^{2}\alpha _{01}^{2}}.
\end{equation*}%
This completes the proof.
\end{proof}

\section{Proof of Corollary \ref{col:cic}}

\begin{proof}
The proof follows once we establish \eqref{asym:index}--\eqref{asym:emp}. 
Our
Condition 7 is the same as \citet[][eq.(2)]{girard2021}. 
Our Condition 2 is sufficient for their second-order Pareto tail condition $\mathcal{C}_{2}\left( \gamma ,\rho ,A\right) $. 
Then \eqref{asym:index} and \eqref{asym:emp} directly follow from their Corollary 2.1. 
Using the same proof of Theorem 4.3.8 in \citet{de2007extreme}, \eqref{asym:quantile} further follows from \eqref{asym:index}, \eqref{asym:emp}, and our Condition 2.
\end{proof}

\end{document}